\documentclass[12pt]{article}

\let\e=\epsilon

\newcommand{\be}{\begin{equation}}
\newcommand{\ee}{\end{equation}}
\newcommand{\bea}{\begin{eqnarray}}
\newcommand{\eea}{\end{eqnarray}}

\newcommand{\del}{\partial}

\begin{document}

\begin{titlepage}
\begin{center}
\vskip .2in \hfill \vbox{
    \halign{#\hfil         \cr
           hep-th/0212145 \cr
           UCI-TR-2002-46\cr
           UCSD-PTH-02-29\cr
           December 2002    \cr
           }  
      }   
\vskip 1.5cm {\Large \bf New Supergravity Solutions for
                          Branes in $AdS_3 \times S^3$  } \\
\vskip .1in \vskip .3in {\bf  Jason Kumar}\footnote{e-mail
address: j1kumar@ucsd.edu}\\
\vskip .15in {\em  Department of Physics,
University of California, San Diego,\\
La Jolla, CA  92093 USA \\} \vskip .2in
{\bf Arvind Rajaraman}\footnote{e-mail
address: arajaram@uci.edu} \vskip .15in {\em
Department of Physics,
University of California, Irvine,  \\
Irvine, CA 92697 USA\\} \vskip .1in \vskip 1cm
\end{center}
\begin{abstract}
We find explicit supergravity solutions which describe branes in
the $AdS_3 \times S^3$ background. These solutions preserve 8 of
the 16 supersymmetries of this background, and are consistent with
$\kappa$-symmetry. These  represent new ${1\over 2}$-BPS states of
string theory. \vskip 0.5cm
\end{abstract}
\end{titlepage}
\newpage

\section{Introduction}

One of the most interesting of the anti-de Sitter solutions of
string theory is the $AdS_3 \times S^3 \times T^4 $ solution which
arises from embedding a stack of fundamental strings within a
stack of NS-5-branes.  This solution is particularly useful
because it is possible to write the action of a string in this
background as a Wess-Zumino-Witten model on the group manifold
$SL(2,R)\times SU(2)$. While the noncompact nature of $SL(2,R)$
leads to subtleties, great progress in understanding the closed
string theory has been made \cite{mo}.  The open string theory on
D-branes in this background has also been studied a great deal
recently \cite{Rajaraman:2001cr}.

It has proven extremely fruitful in the AdS/CFT correspondence to
be able to describe the same system in two different ways; as a
solution to perturbative string theory and as a solution to
classical supergravity. If this can be done, we can generate new
dual descriptions of gravitational theories by gauge theories. We
will therefore try to construct supergravity solutions for
D-branes in $AdS_3$, which will then be described in a dual
description by a field theory. In the particular case that we
shall analyze, the supergravity solution we find is believed to be
dual to a defect conformal field theory on the boundary
\cite{Karch:2000ct}.

There have been several partial results in the previous literature
for such classical supergravity solutions. Supergravity solutions
for general intersecting branes were considered in
\cite{Rajaraman:2002vf,Rajaraman:2000dn,
Rajaraman:2000ws,Fayyazuddin:2000em,Brinne:2000nf,Brinne:2000fh,
Cherkis:2002ir,Lunin:2001jy}.
Secondly, in \cite{Bachas:2000ik},
brane probes were considered.  The equations of motion derived
from the Born-Infeld action were then solved to produce stable
supersymmetric branes. Unfortunately, the latter approach only
produces solutions to linearized supergravity. Since we want to
construct the complete nonlinear solution, we will follow the
general approach of \cite{Rajaraman:2000dn,Fayyazuddin:2000em}.

There is an important feature of these branes that simplifies our
task. It was shown in \cite{Bachas:2000fr}, using boundary state
arguments, that a D-3-brane stretched along an $AdS_2 \times S^2$
submanifold of the background satisfies the equations of motion
and is ${1\over 2}$ BPS; in other words it preserves 8
supercharges. Note that this is twice the number preserved by a
system containing D3-branes, fundamental strings and NS-5 branes.
The point is that in the near horizon limit, the supersymmetry is
enhanced \cite{Kallosh:1997qw}; $AdS_3 \times S^3 \times T^4 $
preserves 16 supercharges. A ${1\over 2}$-BPS brane in this
background therefore preserves 8 supercharges. We therefore expect
considerable simplifications since the branes preserve a greater
amount of supersymmetry.

In the following section, we analyze $\kappa$-symmetry for the
D-branes in $AdS_3 \times S^3$.  This analysis tells us which
Killing spinors are preserved in the presence of the brane.
Specifically, $\kappa$-symmetry informs us that the preserved Killing
spinors are found by applying a particular projector to the
Killing spinors.

We then use this to analyze the Killing equations. We require that
the Killing equations be satisfied once the projector found above
is imposed on the spinors. This then imposes constraints on the
metric and field strengths. We can then solve these constraints to
find the full solution. Since this procedure is rather tedious, we
will simplify by assuming that the axion and dilaton are constant
(we will look at the more general case in future work.)

We find that the constraints can indeed be solved, and the
explicit solution can be found. The sources are found to be
localized at antipodal points on the $S^3$ and wrap an $AdS_2$ in
$AdS_3$.
 This then provides a new ${1\over 2}$-BPS solution of
 supergravity.

\section{Imposing $\kappa$-Symmetry}

The $AdS_3 \times S^3\times T^4$ solution of type IIB supergravity
can be obtained by  taking the near-horizon limit of the solution
generated by  fundamental strings and  NS-5-branes. This system
preserves 16 supersymmetries. We shall ignore the $T^4$ directions
in the subsequent discussions (it is thereby implicitly assumed
that all branes are smeared on the $T^4$).

The metric of $AdS_3 \times S^3$ in global coordinates with unit
normalized radius is \bea ds^2 = d\psi^2 +\cosh^2 \psi
\left(d\omega^2 - \cosh^2 \omega d\tau^2 \right)+~~~~~~~~~~
\nonumber\\
+ d\theta^2 +\sin^2 \theta \left( d\phi^2 + \sin^2 \phi d\chi^2
\right) \eea

It was argued in \cite{Bachas:2000fr} from boundary state
considerations that a D-3-brane could be added in a way which
preserved one half of the supersymmetries. In this embedding, the
geometry of the D3-brane is $AdS_2 \times S^2$.  With a
particularly simple choice of parameters\footnote{$q=0$ in the
notation of \cite{Bachas:2000fr}.}, the D3-brane stretches along
the coordinates $(\tau , \omega, \phi ,\chi)$, with the
coordinates $\psi$ and $\theta$ appearing as  transverse scalars.
We can then find a brane solution by solving the Born-Infeld
equations of motion.

We need to know the background fields. The relevant part of the
$B_{NS}$ background is \bea \overline{B}_{\phi \chi} &=& {1\over
2} \left(\theta -{\sin 2\theta \over 2}\right) \sin \phi
\nonumber\\
\overline{B}_{\tau \omega} &=& {1\over 2} \left(\psi +{\sinh 2\psi
\over 2}\right) \cosh \omega  \eea We can also turn on a magnetic
flux on the D-3-brane of the form \be 4\pi \alpha' F_{\phi \chi }
= -\pi p \sin \phi \ee where $p$ is constant. The Lagrangian for a
D-3-brane embedded in this way is \be L_{DBI} = -T_D \sqrt{-\det
M} \ee where \bea \sqrt{-\det M} &=& N(\psi) L (\theta) \cosh
\omega \sin \phi
\nonumber\\
L(\theta ) &=& \left(\sin^4 \theta + \left(\theta -{\sin 2\theta
\over 2} - {\pi p }\right)^2 \right)^{1\over 2}
\nonumber\\
N(\psi) &=& \left(\cosh^4 \psi -\left(\psi +{\sinh 2\psi \over
2}\right)^2 \right)^{1\over 2} . \eea The solution of the $\theta$
equation of motion is then $\theta_0 = {\pi p } $. Similarly, we
can solve the $\psi$ equation of motion by setting $\psi=0$.

If we set \bea a &=&{1\over 2} \sin {2\pi p } =\sin \theta_0 \cos
\theta_0
\nonumber\\
b &=& \sin^2 {\pi p } =\sin^2 \theta_0  \eea then \bea L(\theta_0)
&=& \sin \theta_0
\nonumber\\
{\cal F}_{\phi \chi} &=& - {1\over 2} \sin 2\theta_0 \sin \phi
\eea

We now discuss the supersymmetries preserved by this brane. For
this we need the $\kappa$-symmetry projector $\Gamma$. This is found
by \cite{Cederwall:1996pv}
\be
d^{p+1} \xi \Gamma = -e^{-\Phi} L_{DBI} ^{-1} e^{\cal F} \wedge
X|_{vol}, \ee where \bea X &=& \bigoplus_n \Gamma_{(2n)} K^n I ,
\nonumber\\
K\psi &=& \psi^*
\nonumber\\
I\psi &=& -\imath \psi
\nonumber\\
\Gamma_{(n)} &=& {1\over n!} d\xi^{i_n} \wedge ... \wedge
d\xi^{i_1} \Gamma_{i_1 ... i_n}  \eea We see that \bea \Gamma &=&
-{1\over \sqrt{a^2 + b^2}} \left(a\gamma_{\tau \omega} KI -
b\gamma_{\tau \omega \phi \chi} K^2 I\right)
\nonumber\\
&=& -\imath (\cos \theta_0 \gamma_{\tau \omega} K + \sin \theta_0
\gamma_{\tau \omega \phi \chi} )  \eea

It may be verified that  $\Gamma$ is traceless and $\Gamma^2 =1$.
This implies that 8 of the Killing spinors (pulled back to the
worldvolume of the brane) will be invariant under the
$\kappa$-symmetry projection $\Gamma\e=\e$.

This is not the end of the story, though. The Killing spinors of
$AdS_3 \times S^3$ in global coordinates can be written as \be
\epsilon = \exp\left({h \psi\over 2} \gamma_{\tau \omega} K\right)
\exp\left({h \theta \over 2} \gamma_{\phi \chi} K\right) R_0 (\phi
,\chi ,\omega ,\tau) \epsilon_0 , \ee where $\epsilon_0$ is an
arbitrary 16-component constant spinor satisfying $\gamma^{2345}
\epsilon_0 = \epsilon_0$, $R_0$ is invertible and where $h=-1$.
Since the $\kappa$-symmetry projection $\Gamma\e=\e$ is imposed on
the brane, the projection can be written explicitly as
\be\label{proj}
 \Gamma \exp\left({h \theta_0 \over 2} \gamma_{\phi \chi} K\right)
  R_0 (\phi ,\chi ,\omega ,\tau) \epsilon_0
=\exp\left({h \theta_0 \over 2} \gamma_{\phi \chi} K\right) R_0
(\phi ,\chi ,\omega ,\tau) \epsilon_0 \ee  Since we want to find
the projection on the full spinor $\e$, we  still need to
conjugate by the ($\psi$,$\theta$) dependence of the Killing
spinor to find the invariant Killing spinor throughout the space.

 Defining \be \Lambda =\exp\left({\psi_0\over 2}
\gamma_{\tau \omega } K\right) \exp\left({\theta_0\over2}
\gamma_{\phi \chi} K\right) \exp\left(-{\psi\over 2} \gamma_{\tau
\omega } K\right) \exp\left(-{\theta\over 2} \gamma_{\phi \chi}
K\right), \ee we see we can rewrite the above projection
(\ref{proj}) as $\tilde{\Gamma}\e=\e$ where\footnote{$\delta
\theta = \theta - \theta_0$. To connect to more general
embeddings, we also write $\delta \psi = \psi - \psi_0$, although
in this case $\psi_0 =0$.} \bea \tilde \Gamma \equiv \Lambda^{-1}
\Gamma \Lambda &=& - {\imath \over \sqrt{a^2 +b^2}} \left(\cos
\delta \theta + \sin \delta \theta  \gamma_{\phi \chi} K\right)
\nonumber\\
&\times &\left(\cosh \delta \psi +\sinh \delta \psi \gamma_{\tau
\omega } K\right) \gamma_{\tau \omega} (aK + b\gamma_{\phi \chi})
  \eea We then have

\bea (1+\tilde \Gamma)  \epsilon &=& [1 - \imath  (M +
N\gamma_{\tau \omega} K +O\gamma_{\phi \chi}K +P\gamma_{\tau
\omega \phi \chi}) ]\epsilon
\nonumber\\
M &=& {1\over \sqrt{a^2 + b^2}} (a\cos \delta \theta\sinh \delta
\psi - b\sin \delta \theta\sinh \delta \psi )
\nonumber\\
N &=& {1\over \sqrt{a^2 + b^2}} (a\cos \delta \theta\cosh \delta
\psi - b\sin \delta \theta\cosh \delta \psi )
\nonumber\\
O &=& {1\over \sqrt{a^2 + b^2}} (a\sin \delta \theta\sinh \delta
\psi + b\cos \delta \theta\sinh \delta \psi )
\nonumber\\
P &=& {1\over \sqrt{a^2 + b^2}} (a\sin \delta \theta\cosh \delta
\psi + b\cos \delta \theta\cosh \delta \psi ) \eea which reduces
to \bea M&=&\sinh\psi\cos\theta\qquad\qquad N=\cosh\psi\cos\theta
\nonumber\\
O&=&\sinh\psi\sin\theta\qquad\qquad P=\cosh\psi\sin\theta  \eea
Note that the $\theta_0$ dependence of the space-time projector
has dropped out.  For any location $\theta_0$ of the brane, the
Killing spinors preserved by $\kappa$-symmetry are the same.

Using this projection and Hodge duality, one can rewrite the
projection as \bea \label{ABdef} \gamma_{\tau \omega} K\epsilon
&=& ({A} +{B} \gamma_{\psi \theta}) \epsilon \eea where \bea
A=-{\cosh\psi\over
(\cosh^2\psi-\sin^2\theta)}(\sinh\psi+i\cos\theta)
\nonumber\\
B={i\sin\theta\over
(\cosh^2\psi-\sin^2\theta)}(\sinh\psi+i\cos\theta) \eea

\section{The Killing equations}

We will impose the projection ${1\over 2} \left(1+\tilde
\Gamma\right) \epsilon =\epsilon$ and demand that the spinors
which satisfy it are solutions of the Killing equations. This will
generate a ${1\over 2}$-BPS solution.\footnote{We assume that no
field strengths have indices along the $T^4$ unless they have
indices in all those directions, and can thus be related to a
field with no such indices by Hodge duality.} Note that because
the choice of Killing spinor is independent of $\theta_0$ (and
presumably $\psi_0$), the solution may correspond to a smeared
brane, and not one necessarily localized at the place originally
anticipated.

The Killing equations are of the form
\bea \del_{\tilde \mu}\e -{1\over 4} \omega_{\tilde \mu} ^{ab}
\gamma_{ab} +{\imath \over 192}  F_{\tilde \mu}^{~bcde}
\gamma_{bcde}\epsilon^* -{\imath \over 48} e^{\Phi} (G^{abc}
\gamma_{\tilde \mu abc}- 9G_{\tilde \mu}^{~ab} \gamma_{ab})
\epsilon^*=0 \eea

 For example, the $\psi$ Killing equation is
\footnote{$\tilde \mu$ is a curved space-time index, while $\mu$
is a tangent-space index. We use the notation $\omega_{\mu} ^{\nu
\gamma} = e_{\mu} ^{\tilde \mu} \omega_{\tilde \mu} ^{\nu
\gamma}$.} \bea {\del_\psi f\over f}\gamma_\psi \epsilon -{1\over
2} \omega_{ \psi} ^{\psi \theta} \gamma_{ \theta} \epsilon
-{\imath \over 8}  F^{\tau\omega \phi \chi \theta}
\gamma_{\tau\omega \phi \chi \theta} \epsilon +{\imath \over 8}
F^{\psi\tau \omega \phi \chi } \gamma_{\psi\tau\omega \phi \chi }
\epsilon~~~~~~~~~~~~~~~
\nonumber\\
+3{\imath \over 8} e^{\Phi} G^{\psi\tau \omega} \gamma_{\psi\tau
\omega}
 \epsilon^*
- {\imath \over 8} e^{\Phi} G ^{\tau\omega \theta}
 \gamma_{\tau\omega \theta}
\epsilon^* +3{\imath \over 8} e^{\Phi} G^{\psi\phi \chi}
\gamma_{\psi\phi \chi}  \epsilon^*
\nonumber\\
-{\imath \over 8} e^{\Phi} G^{\phi \chi \theta} \gamma_{\phi \chi
\theta} \epsilon^*
=
 \nonumber\\
 {e_{\psi} ^{\tilde \psi}
 }\left(-{1\over 2}
\bar{\omega}_{ \psi} ^{\psi \theta} \gamma_{ \theta} \epsilon
+3{\imath \over 8} e^{\Phi} \bar{G} ^{\psi\tau \omega}
\gamma_{\psi\tau \omega}
 \epsilon^*
 -{\imath \over 8} e^{\Phi} \bar{G}^{\phi \chi \theta} \gamma_{\phi
\chi \theta} \epsilon^*\right)   \eea where the bars over
expressions indicate  that  the expression is to be evaluated in
the unperturbed background $AdS$ solution and where we have made
the ansatz $\epsilon = f(\psi ,\theta) \overline{\epsilon}$.
The other Killing equations are entirely similar.

The axion-dilaton equation is

\bea \label{axioneq}
 {1\over 4}\del_\psi \Phi \gamma_{ \psi}\epsilon
 +{1\over 4}\del_\theta \Phi\gamma_{\theta}\epsilon+{\imath \over 8}
e^{\Phi} G ^{\omega\psi \tau} \gamma_{\omega\psi \tau}
 \epsilon^*
+ {\imath \over 8} e^{\Phi} G ^{ \omega\tau \theta}
 \gamma_{\omega\tau \theta}
\epsilon^* \nonumber\\
+{\imath \over 8} e^{\Phi} G^{\psi \phi \chi} \gamma_{\psi \phi
\chi} \epsilon^* +{\imath \over 8} e^{\Phi} G^{\phi \chi \theta}
\gamma_{\phi \chi \theta} \epsilon^*=
 \nonumber\\
{\imath \over 8} e^{\Phi} \overline{G ^{\omega\psi \tau}}
\gamma_{\omega\psi \tau} \epsilon^* +{\imath \over 8} e^{\Phi}
\overline{G^{\phi \chi \theta}} \gamma_{\phi \chi \theta}
\epsilon^* \eea

We shall now assume that $\Phi = 0$. Then  (\ref{axioneq}) is
solved by the ansatz \bea \Phi &=& 0
\nonumber\\
G^{\psi \tau \omega } &=& G^{\phi \chi \theta}
\nonumber\\
G^{\tau \omega \theta } &=& -G^{\psi \phi \chi}  \eea

We can now return to the other Killing equations. In each of them,
we first impose the projector (\ref{ABdef}). The Killing equation
then becomes a matrix equation which is to be satisfied
identically. Thus the complex coefficient of each matrix must be
zero. In this way, each Killing equation leads to two complex
algebraic equations.

For example, the  $\psi$ Killing equation, after this procedure,
leads to the equations
 \bea
 {\del_\psi f\over f}
 +{\imath \over 8}  F^{\tau\omega \phi \chi
\theta}  +{\imath \over 8} e^{\Phi} \left(3G ^{\psi\tau \omega} {
A} +  G ^{\tau\omega \theta}
 { B}
 -{3}  G^{\psi\phi \chi}
{ B} +G^{\phi \chi \theta} { A}\right)  =
 \nonumber\\
 +  {3\imath \over 8} e^{\Phi}
\bar{G}^{\psi\tau \omega} {A}
 +{\imath \over 8} e^{\Phi} \bar{G}^{\phi \chi \theta}
 { A}
\eea

and

 \bea
 -
{1\over 2} \omega_{ \psi} ^{\psi \theta}  +{\imath \over 8}
 F^{\psi\tau \omega \phi \chi } +{\imath \over 8}
e^{\Phi} \left(3G ^{\psi\tau \omega} {B}  - G ^{\tau\omega \theta}
 { A}
 +{3}  G^{\psi\phi \chi}
{ A}  + G^{\phi \chi \theta} { B}\right) =
 \nonumber\\
 -{1\over 2} \bar{\omega}_{ \psi} ^{\psi
\theta}   +3{\imath \over 8} e^{\Phi} \bar{G}^{\psi\tau \omega} {
B}
 +{\imath \over 8} e^{\Phi} \bar{G}^{\phi \chi \theta}
 { B}
\eea

By taking linear combinations of these equations such that the
barred field strengths cancel,  we obtain the equations \bea
\label{STeq} {e_{\tilde \theta} ^{\theta} \over
\overline{e}_{\tilde \theta} ^{\theta} } {e_{\tilde \phi} ^{\phi}
\over \overline{e}_{\tilde \phi} ^{\phi} } =  S(\theta)
\qquad\qquad {e_{\tilde \psi} ^{\psi} \over \overline{e}_{\tilde
\psi} ^{\psi} } {e_{\tilde \omega} ^{\omega} \over
\overline{e}_{\tilde \omega} ^{\omega} } =  T(\psi)  \eea and
 \bea
2{\partial_{\psi} f\over f} + \omega_{\phi} ^{\phi \psi} =
2{\partial_{\theta} f\over f} + \omega_{\omega} ^{\omega \theta}
=0 , \eea which tells us that \be  \overline{e_{\tilde \omega}
^{\omega}}{e_{\tilde \phi} ^{\phi}
 } = \overline{e_{\tilde \phi} ^{\phi}}{e_{\tilde \omega}
^{\omega}  }  \ee

One can substitute all of this back into the Killing equations to
derive further relations involving the barred quantities, such as

\bea {\imath \over 8} e^{\Phi} (3 \overline{G^{\psi \tau \omega} }
+\overline{G^{\phi \chi \theta}}  ) \left({e^{\tilde \psi} _{\psi}
\over \overline{e}^{\tilde \psi} _{\psi} } - {e^{\tilde \omega}
_{\omega} \over \overline{e}^{\tilde \omega} _{\omega} }\right) A
&=& {\partial_{\psi} f\over f} +{1\over 2} \omega_{\omega}
^{\omega \psi} -{1\over 2} \overline{\omega_{\omega} ^{\omega
\psi}} {e_{\omega } ^{\tilde \omega} \over \overline{e}_{\omega }
^{\tilde \omega} }
\nonumber\\
+ {\imath \over 4} F^{\tau \omega \phi \chi \theta}
&+& {\imath \over 2} e^{\Phi} B(G^{\tau \omega \theta} - G^{\psi
\phi \chi}) . \eea

These relations allow us to determine the entire solution,
including the fact that $S(\theta) = T(\psi) =const$.

Defining $\gamma = \sin \theta \cosh \psi = \beta \cosh^2 \psi$,
the complete solution for the metric is found to be\bea ds^2 =
H^{1\over 2} d\theta^2 &+& H^{-{1\over 2}}\sin^2 \theta ( d\phi^2
+ \sin^2 \phi d\chi^2 )
\nonumber\\
&+&H^{1\over 2} d\psi^2 + H^{-{1\over 2}}\cosh^2 \psi (d\omega^2 -
\cosh^2 \omega d\tau^2 ) ,
\nonumber\\
H&=&({c\gamma \over 1+c\gamma})^2 . \eea and the field strengths
are \bea
F^{\tau \omega \phi \chi \theta} &=& F^{\psi \tau \omega \phi \chi
} =0 ,
\nonumber\\
G^{\tau \omega \theta} =-G^{\psi \phi \chi } &=& {1\over 2}
{1\over \sqrt {c\gamma(1+c\gamma)}} \left({\beta \over \sin^2
\theta} +\imath \cot \theta \tanh \psi \right)
\nonumber\\
G^{\psi \tau \omega} = G^{\phi \chi \theta} &=& -\imath
\sqrt{1+c\gamma \over c\gamma } \left(1 - {1\over 2} {1\over
1+c\gamma}\right) \eea where $c$ is a constant of integration.

One may verify that
the Bianchi identity $dG=0$ is satisfied away from $\theta=0$.  We
may thus write $G=dC$, where

\bea C_{\tilde \tau \tilde \omega}&=& {\imath \over 2} \left(\psi
+{\sinh 2\psi \over 2 } +{\sinh \psi \cosh \omega \over c\sin
\theta }\right) +{\cosh \omega \over 2c} \cot \theta
\nonumber\\
C_{\tilde \phi \tilde \chi}&=& {\imath \over 2} \left(\theta
-{\sin 2\theta \over 2 } +{\cos \theta \sin \phi \over c\cosh \psi
}\right) -{\sin \phi \over 2c} \tanh \psi \eea

The gauge fields are
singular at $\theta =0,\pi$, indicating a source there.
The net D1-brane charge may then be found up to a normalization
constant to be \be Q = Re \int d\psi d\phi d\chi G_{\tilde \psi
\tilde \phi \tilde \chi} = {2\pi \over c} \int d\psi {1\over
\cosh^2 \psi} ={4\pi \over c}\ee
 It is also easily seen that the 3-brane charge is zero.
The sources can thus be interpreted as string-like objects
wrapping the ($\tau$,$\omega$) directions in $AdS_3$, and sitting at
antipodal points $\theta=0,\pi$ on the $S^3$.

To conclude, we have found a new solution of  supergravity,
representing a brane in an $AdS_3 \times S^3$ background, which
preserves 8 supersymmetries.
 The preserved killing spinor is independent of the embedding
parameters $\theta_0 ,\psi_0$, demonstrating that the sources can
be smeared or superposed without breaking additional
supersymmetry.

The specific case we have analyzed has no 3-brane charge. It is
likely that generalizing our ansatz to more general cases with
nonconstant scalar fields will produce solutions involving
three-branes as well. We will explore this possibility in future
work \cite{future}. It will also be interesting to analyze the
implications for the dual defect conformal field theory.

\vskip .2in {\bf Acknowledgements}

We thank K. Intriligator for helpful discussions. J. K. is
supported by DOE-FG03-97ER40546.  A. R. gratefully acknowledges
the hospitality of the Aspen Center for Physics, where part of
this work was done.

\section{Appendix:  Notation}

In \cite{Bergshoeff:1996wk} $\epsilon$ is a 32-component complex
spinor in mostly - signature, whereas \cite{Maldacena:1996ky} uses
$\epsilon_{L,R}$, each of which is a 16-component Majorana spinor
in mostly + signature.

Considering the case of Type IIB, if we choose the $\gamma$'s to
be real and define \bea \epsilon_{R} &=& Re ({1 \pm \gamma^{11}
\over 2} \epsilon)
\nonumber\\
\epsilon_{L} &=& Im ({1 \pm \gamma^{11} \over 2} \epsilon)
\nonumber\\
\gamma^{\mu} &=& \imath \Gamma^{\mu} \eea then the conventions are
consistent.

We use the notation of \cite{Bergshoeff:1996wk}, but will switch
to mostly + signature.  In the brane action, we use the same sign
convention as \cite{Kallosh:1997jz} .


\begin{thebibliography}{99}

\bibitem{mo}
J.~M.~Maldacena and H.~Ooguri,
Phys.\ Rev.\ D {\bf 65}, 106006 (2002) [arXiv:hep-th/0111180].
\\
J.~M.~Maldacena, H.~Ooguri and J.~Son,
J.\ Math.\ Phys.\  {\bf 42}, 2961 (2001) [arXiv:hep-th/0005183].
\\
J.~M.~Maldacena and H.~Ooguri,
J.\ Math.\ Phys.\  {\bf 42}, 2929 (2001) [arXiv:hep-th/0001053].

\bibitem{Rajaraman:2001cr}
A.~Rajaraman and M.~Rozali, ``Boundary states for D-branes in
AdS(3),'' Phys.\ Rev.\ D {\bf 66}, 026006 (2002)
[arXiv:hep-th/0108001].
\\
B.~Ponsot, V.~Schomerus and J.~Teschner, ``Branes in the Euclidean
AdS(3),'' JHEP {\bf 0202}, 016 (2002) [arXiv:hep-th/0112198].
\\
P.~Lee, H.~Ooguri and J.~w.~Park, ``Boundary states for AdS(2)
branes in AdS(3),'' Nucl.\ Phys.\ B {\bf 632}, 283 (2002)
[arXiv:hep-th/0112188].
\\
P.~Lee, H.~Ooguri, J.~w.~Park and J.~Tannenhauser, ``Open strings
on AdS(2) branes,'' Nucl.\ Phys.\ B {\bf 610}, 3 (2001)
[arXiv:hep-th/0106129].
\\
A.~Rajaraman, ``New AdS(3) branes and boundary states,''
arXiv:hep-th/0109200.
\\
A.~Giveon, D.~Kutasov and A.~Schwimmer, ``Comments on D-branes in
AdS(3),'' Nucl.\ Phys.\ B {\bf 615}, 133 (2001)
[arXiv:hep-th/0106005].
\\
T.~Quella,
``On the hierarchy of symmetry breaking D-branes in group manifolds,''
JHEP {\bf 0212}, 009 (2002)
[arXiv:hep-th/0209157].
\\
S.~Ribault,
``Two AdS(2) branes in the Euclidean AdS(3),''
arXiv:hep-th/0210248.



\bibitem{Karch:2000ct}
A.~Karch and L.~Randall, ``Locally localized gravity,'' JHEP {\bf
0105}, 008 (2001) [arXiv:hep-th/0011156].
\\
A.~Karch and L.~Randall, ``Open and closed string interpretation
of SUSY CFT's on branes with  boundaries,'' JHEP {\bf 0106}, 063
(2001) [arXiv:hep-th/0105132].
\\
O.~DeWolfe, D.~Z.~Freedman and H.~Ooguri, ``Holography and defect
conformal field theories,'' Phys.\ Rev.\ D {\bf 66}, 025009 (2002)
[arXiv:hep-th/0111135].
\\
C.~Bachas, J.~de Boer, R.~Dijkgraaf and H.~Ooguri, ``Permeable
conformal walls and holography,'' JHEP {\bf 0206}, 027 (2002)
[arXiv:hep-th/0111210].
\\
J.~Erdmenger, Z.~Guralnik and I.~Kirsch, ``Four-dimensional
superconformal theories with interacting boundaries or  defects,''
Phys.\ Rev.\ D {\bf 66}, 025020 (2002) [arXiv:hep-th/0203020].
\\
N.~R.~Constable, J.~Erdmenger, Z.~Guralnik and I.~Kirsch,
``Intersecting D3-branes and holography,''
arXiv:hep-th/0211222.
\\
N.~R.~Constable, J.~Erdmenger, Z.~Guralnik and I.~Kirsch,
``(De)constructing intersecting M5-branes,''
arXiv:hep-th/0212136.




\bibitem{Rajaraman:2002vf}
A.~Rajaraman, ``Comments on D-branes in flux backgrounds,''
arXiv:hep-th/0208085.


\bibitem{Rajaraman:2000dn}
A.~Rajaraman, ``Supergravity duals for N = 2 gauge theories,''
JHEP {\bf 0210}, 009 (2002) [arXiv:hep-th/0011279].

\bibitem{Rajaraman:2000ws}
A.~Rajaraman, ``Supergravity solutions for localised brane
intersections,'' JHEP {\bf 0109}, 018 (2001)
[arXiv:hep-th/0007241].

\bibitem{Fayyazuddin:2000em}
A.~Fayyazuddin and D.~J.~Smith, ``Warped AdS near-horizon geometry
of completely localized intersections  of M5-branes,'' JHEP {\bf
0010}, 023 (2000) [arXiv:hep-th/0006060].

\bibitem{Brinne:2000nf}
B.~Brinne, A.~Fayyazuddin, T.~Z.~Husain and D.~J.~Smith, ``N = 1
M5-brane geometries,'' JHEP {\bf 0103}, 052 (2001)
[arXiv:hep-th/0012194].



\bibitem{Brinne:2000fh}
B.~Brinne, A.~Fayyazuddin, S.~Mukhopadhyay and D.~J.~Smith,
``Supergravity M5-branes wrapped on Riemann surfaces and their QFT
duals,'' JHEP {\bf 0012}, 013 (2000) [arXiv:hep-th/0009047].

\bibitem{Cherkis:2002ir}
S.~A.~Cherkis and A.~Hashimoto, ``Supergravity solution of
intersecting branes and AdS/CFT with flavor,''
arXiv:hep-th/0210105.

\bibitem{Lunin:2001jy}
O.~Lunin and S.~D.~Mathur,
``AdS/CFT duality and the black hole information paradox,''
Nucl.\ Phys.\ B {\bf 623}, 342 (2002)
[arXiv:hep-th/0109154].
\\
O.~Lunin, S.~D.~Mathur and A.~Saxena,
``What is the gravity dual of a chiral primary?,''
arXiv:hep-th/0211292.





\bibitem{Bachas:2000ik}
C.~Bachas, M.~R.~Douglas and C.~Schweigert, ``Flux stabilization
of D-branes,'' JHEP {\bf 0005}, 048 (2000) [arXiv:hep-th/0003037].
\\
A.~Alekseev and V.~Schomerus, ``RR charges of D2-branes in the WZW
model,'' arXiv:hep-th/0007096.
\\
A.~Y.~Alekseev, A.~Recknagel and V.~Schomerus, ``Brane dynamics in
background fluxes and non-commutative geometry,'' JHEP {\bf 0005},
010 (2000) [arXiv:hep-th/0003187].
A.~Y.~Alekseev, S.~Fredenhagen, T.~Quella and V.~Schomerus,
``Non-commutative gauge theory of twisted D-branes,''
Nucl.\ Phys.\ B {\bf 646}, 127 (2002)
[arXiv:hep-th/0205123].



\bibitem{Bachas:2000fr}
C.~Bachas and M.~Petropoulos, ``Anti-de-Sitter D-branes,'' JHEP
{\bf 0102}, 025 (2001) [arXiv:hep-th/0012234].

\bibitem{Kallosh:1997qw}
R.~Kallosh and J.~Kumar, ``Supersymmetry enhancement of D-p-branes
and M-branes,'' Phys.\ Rev.\ D {\bf 56}, 4934 (1997)
[arXiv:hep-th/9704189].




\bibitem{Cederwall:1996pv}
M.~Cederwall, A.~von Gussich, B.~E.~Nilsson and A.~Westerberg,
``The Dirichlet super-three-brane in ten-dimensional type IIB
supergravity,'' Nucl.\ Phys.\ B {\bf 490}, 163 (1997)
[arXiv:hep-th/9610148].
\\
M.~Aganagic, C.~Popescu and J.~H.~Schwarz, ``D-brane actions with
local kappa symmetry,'' Phys.\ Lett.\ B {\bf 393}, 311 (1997)
[arXiv:hep-th/9610249].
\\
M.~Cederwall, A.~von Gussich, B.~E.~Nilsson, P.~Sundell and
A.~Westerberg, ``The Dirichlet super-p-branes in ten-dimensional
type IIA and IIB  supergravity,'' Nucl.\ Phys.\ B {\bf 490}, 179
(1997) [arXiv:hep-th/9611159].
\\
E.~Bergshoeff and P.~K.~Townsend, ``Super D-branes,'' Nucl.\
Phys.\ B {\bf 490}, 145 (1997) [arXiv:hep-th/9611173].
\\
M.~Aganagic, C.~Popescu and J.~H.~Schwarz, ``Gauge-invariant and
gauge-fixed D-brane actions,'' Nucl.\ Phys.\ B {\bf 495}, 99
(1997) [arXiv:hep-th/9612080].
\\
E.~Bergshoeff, R.~Kallosh, T.~Ortin and G.~Papadopoulos,
``kappa-symmetry, supersymmetry and intersecting branes,'' Nucl.\
Phys.\ B {\bf 502}, 149 (1997) [arXiv:hep-th/9705040].
\\
K.~Skenderis and M.~Taylor,
JHEP {\bf 0206}, 025 (2002) [arXiv:hep-th/0204054].

\bibitem{future}
J.~Kumar and A.~Rajaraman, work in progress.

\bibitem{Bergshoeff:1996wk}
E.~Bergshoeff, ``p-branes, D-branes and M-branes,''
arXiv:hep-th/9611099.

\bibitem{Maldacena:1996ky}
J.~M.~Maldacena, ``Black holes in string theory,''
arXiv:hep-th/9607235.







\bibitem{Kallosh:1997jz}
R.~Kallosh, J.~Kumar and A.~Rajaraman, ``Special conformal
symmetry of worldvolume actions,'' Phys.\ Rev.\ D {\bf 57}, 6452
(1998) [arXiv:hep-th/9712073].



\end{thebibliography}
\end{document}